\documentclass[%
 twocolumn,10pt,superscriptaddress,longbibliography,amsmath,amssymb,aps,pra,
]{revtex4-2}

\usepackage{graphicx}% Include figure files
\usepackage{dcolumn}% Align table columns on decimal point
\usepackage{bm}% bold math
\usepackage{qcircuit}
\usepackage{epstopdf}
\usepackage[caption=false]{subfig}
\usepackage{color}
\usepackage{float}
\usepackage{url}
\usepackage[colorlinks=true,linkcolor=blue,citecolor=red, linktocpage=true,breaklinks=true]{hyperref}

\newcommand{\tr}{\mathrm{Tr}}
\begin{document}

%\title{Quantum-search-oriented benchmarks for quantum computers}% Force line breaks with \\
\title{Quantum Search on Noisy Intermediate-Scale Quantum Devices}% \\

\author{Kun \surname{Zhang}}
\affiliation{Department of Chemistry, Stony Brook university, Stony Brook, New York 11794, USA}

\author{Kwangmin \surname{Yu}}
\affiliation{Computational Science Initiative, Brookhaven National Laboratory, Upton, New York 11973, USA}

\author{Vladimir \surname{Korepin}}
%\email{Email: vladimir.korepin@stonybrook.edu}
\affiliation{C.N. Yang Institute for Theoretical Physics, Stony Brook university, Stony Brook, New York 11794-3840, USA}

\date{\today}

\begin{abstract}

    Quantum search algorithm (also known as Grover's algorithm) lays the foundation for many other quantum algorithms. Although it is very simple, its implementation is limited on noisy intermediate-scale quantum (NISQ) processors. Grover's algorithm was designed without considering the physical resources, such as depth, in the real implementations. Therefore, Grover's algorithm can be improved for NISQ devices. In this paper, we demonstrate how to implement quantum search algorithms better on NISQ devices. We present detailed benchmarks of the five-qubit quantum search algorithm on different quantum processors, including IBMQ, IonQ, and Honeywell quantum devices. We report the highest success probability of the five-qubit search algorithm compared to previous works. Our results show that designing the error-aware quantum search algorithms is possible, which can maximally harness the power of NISQ computers.
	
\end{abstract}

\maketitle

\section{\label{sec:intro}Introduction}

The ultimate goal of quantum computers is to implement quantum algorithms that are superior to the classical counterpart. During the last twenty years, quantum computers have vastly developed. Quantum processors with hundred qubits have been delicately designed and built. We have entered the noisy-intermediate-quantum (NISQ) era \cite{preskillQuantumComputingNISQ2018}. Quantum processors with hundred qubits have the ability to tackle problems far beyond the reach of classical computers. However, errors limit the number of consecutive operations that can be applied. The number of consecutive operations is also called depth. 

To harness the power of NISQ processors, various quantum algorithms with shallow depths have been designed \cite{bhartiNoisyIntermediatescaleQuantum2022}. The practical power of the shallow depth algorithm is under extensive study now. Nevertheless, the promising of quantum computers largely relies on the application of functional quantum algorithms, such as Shor's algorithm \cite{shorAlgorithmsQuantumComputation1994} and Grover's algorithm \cite{groverQuantumMechanicsHelps1997}. Recently, more researches begin to benchmark the quantum computers based on those applications (application-oriented benchmarking) \cite{lubinskiApplicationOrientedPerformanceBenchmarks2021,georgopoulosQuantumComputerBenchmarking2021}. 

Because of its simplicity, Grover's algorithm is usually the first quantum algorithm taught in the course of quantum computation. Grover's algorithm solves the unstructured search problem \cite{groverQuantumMechanicsHelps1997,giriReviewQuantumSearch2017}. It finds the target item, also called the marked item, in an unstructured database. Classically, the unstructured search problem can be solved by querying each item in the database. The target item can be recognized by the black-box function (oracle). Therefore, the oracular complexity of the classical search is $\mathcal O(N)$, assuming that the number of items in the database is $N$. Grover's algorithm has the oracular complexity $\mathcal O(\sqrt N)$. The quadratic speedup is due to the superposition of quantum states. 

Grover's algorithm is optimal in the number of queries to the oracle, because of the linearity of quantum mechanics \cite{boyerTightBoundsQuantum1998a,zalkaGroverQuantumSearching1999}. The idea of Grover's algorithm is not limited to the unstructured search problem. The amplitude of wanted states can be amplified with a similar quadratic speedup \cite{brassardQuantumAmplitudeAmplification2002,groverQuantumComputersCan1998}. Therefore, it makes the generalized quantum search algorithm applicable to a wide of problems, such as quantum machine learning \cite{biamonteQuantumMachineLearning2017a}. 

During the last twenty years, many variants of Grover's algorithm have been proposed \cite{galindoFamilyGroverQuantumsearching2000,long2001grover,groverTradeoffsQuantumSearch2002,katoGroveralgorithmlikeOperatorUsing2005,tulsiGeneralFrameworkQuantum2012,tulsiFasterQuantumSearching2015,jiangNearoptimalQuantumCircuit2017,gilliamOptimizingQuantumSearch2020,zhangDepthOptimizationQuantum2020a,brianskiIntroducingStructureExpedite2021,liuHardwareEfficientQuantum2021,kwonQuantumAmplitudeamplificationOperators2021}. Few of them have focused on the practical performance of quantum search algorithms on NISQ devices. It was Grover himself who firstly realized the trade-off between the number of oracles and the physical resource for real implementations (such as depths) \cite{groverTradeoffsQuantumSearch2002}. And then from a more practical viewpoint, one can variational learning the physical resource of quantum search algorithms by comparing all the possible implementations \cite{zhangDepthOptimizationQuantum2020a}. Besides, the divide-and-conquer strategy can also be applied to the quantum search algorithm, which prevents the error accumulations \cite{zhangDepthOptimizationQuantum2020a}. Note that the depth optimization of quantum search algorithms can also benefit their implementations in the post NISQ era. In other words, reducing the circuit depth also reduces the error-correction resources (and the running time). 

The implementation of Grover's algorithm (with the three-qubit search domain) was firstly reported in 2017 \cite{figgattComplete3QubitGrover2017}. Since then, more researches have studied the performance of Grover's algorithm on real quantum processors \cite{mandviwallaImplementingGroverAlgorithm2018,satohSubdividedPhaseOracle2020,gwinnerBenchmarking16elementQuantum2021,hlembotskyiEfficientUnstructuredSearch2020,zhangImplementationEfficientQuantum2021}. Most of the realizations are up to four qubits ($N=2^4$). 

In our study, we benchmark the state-of-art quantum processors via the quantum search algorithms. The significance of benchmark in our work is three-fold. First, the recent application-oriented benchmark reported in \cite{lubinskiApplicationOrientedPerformanceBenchmarks2021,georgopoulosQuantumComputerBenchmarking2021} gives a general framework on how to benchmark quantum computers via various quantum algorithms. However, a more sophisticated design of the circuits can improve the performance. For example, being aware of the connectivity of physical qubits can reduce unnecessary SWAP gates in the circuit. One can also take the advantage of the relative-phase Toffoli gates to reduce the circuit depth \cite{songSimplifiedToffoliGate2003,maslovAdvantagesUsingRelativephase2016}. We apply the above two techniques to the circuit design and focus on the performance of \textit{five-qubit} quantum search algorithm ($N=2^5$). We achieve the highest success probability of the five-qubit search algorithm compared to previous works \cite{hlembotskyiEfficientUnstructuredSearch2020,zhangImplementationEfficientQuantum2021}. Second, cloud quantum computations, based on different types of quantum computers, are available to researchers, such as IBMQ \cite{IBM}, IonQ \cite{IonQ} and Honeywell quantum systems (Honeywell Quantum Solution combines Cambridge Quantum called Quantinuum) \cite{Quantinuum}. We aim to benchmark the same algorithms across the different quantum processors. The results can shed light on different aspects of quantum processors realized by different physical systems. Third, we include the depth-optimized and divide-and-conquer quantum search algorithms in our benchmark. The results can demonstrate the optimal way to implement quantum search algorithms on NISQ devices. 

The paper is organized as follows. We introduce Grover's algorithm, depth-optimized and divide-and-conquer quantum search algorithms in Sec. \ref{sec:QSA}. Then we present the benchmark metrics, circuits, results, and the noisy simulation benchmarks in \ref{sec:benchmark_results}. Sec. \ref{sec:conclusion} is the conclusions and outlooks. The realization of five-qubit Toffoli gate and the circuit setups are presented in Appendices \ref{App:circuits} and \ref{App:setup}.

\section{\label{sec:QSA}Quantum search algorithms}

In this section, we give a brief review of different realizations of quantum search algorithms. The original Grover's algorithm, the depth-optimized quantum search algorithm, and the divide-and-conquer quantum search algorithm are introduced in Sec. \ref{subsec:Grover}, \ref{subsec:depth_opt} and \ref{subsec:divde_conquery}.

\subsection{\label{subsec:Grover}Grover's algorithm}

Suppose that the number of items in the database, denoted as $N$, is a power of 2. Then there requires $n=\log_2N$ number of qubits to represent all the items. The index of each item corresponds to one basis vector $|j\rangle$ in $\mathcal H_2^{\otimes n}$. For convenience, the basis vector is chosen as the computational basis (bit strings of zeros and ones). The initial state is set as the uniform superpositions of all the basis vectors, denoted as $|s_n\rangle$. Although such a quantum state is highly nontrivial, it can be easily prepared with a one-depth circuit, given by
\begin{equation}
    |s_n\rangle = H^{\otimes}|0\rangle^{\otimes}.
\end{equation}
Here $H$ is the single-qubit Hadamard gate \cite{NC10}. 

The information of the target state or the target item is encoded in the oracle. The oracle, also called the black box, can distinguish the target and non-target state. In quantum search algorithms, querying the oracle (implementing the oracle) would reflect the sign of target state amplitude. The corresponding mathematical expression is 
\begin{equation}
    \mathcal O_t = 1\!\!1_{2^n} - 2|t\rangle\langle t|,
\end{equation}
with the target state denoted as $|t\rangle$. Here we assume the uniqueness of the target state for simplicity. Grover's algorithm also works if there are multiple target states \cite{boyerTightBoundsQuantum1998a}.

The oracle $\mathcal O_t$ is not enough to pick up the target state from $|s_n\rangle$. Define the operator $D_n$ as
\begin{equation}
    D_n = 1\!\!1_{2^n} - 2|s_n\rangle\langle s_n|,
\end{equation}
which is also called the diffusion operator. The operator $D_n$ reflects all amplitudes in terms of their average. Intuitively, the operator $D_n$ ``diffuses'' the amplitudes of target state with the uniformed amplitudes of non-target states. 

Grover's algorithm is realized by repeatedly applying the operator (also called the Grover operator)
\begin{equation}
    G_n = D_n O_t
\end{equation}
on the initial state $|s_n\rangle$. Then the amplitude of the target state increases in a {\it nonlinear} way. It gives
\begin{equation}
    |\langle t|G^j_n|s_n\rangle|^2 = \sin^2((2j+1)\theta),
\end{equation}
with integer $j$ and $\theta = \arcsin(1/\sqrt N)$. When the number of iteration $j$ approaches $\pi\sqrt N/4$, the success probability of finding the target state approaches $1$. We can conclude that the oracular complexity of Grover's algorithm is $\mathcal O(\sqrt N)$. It is a quadratic speedup compared to the classical complexity $\mathcal O(N)$. 

The increasing speed of target amplitude is nonlinear in Grover's algorithm. When the number of iteration $j$ approaches $\pi\sqrt N/4$, the algorithm becomes less efficient. It has been argued in \cite{boyerTightBoundsQuantum1998a,gingrichGeneralizedQuantumSearch2000} that Grover's algorithm stops at $0.5829\sqrt N$ is the most efficient way. The corresponding success probability is around $0.8446$. In Sec. \ref{subsec:depth_opt}, we will show that the above strategy is also the optimal way to implement Grover's algorithm when optimizing the depth resource.

\subsection{\label{subsec:depth_opt}Depth-optimized quantum search algorithms}

Different unstructured search problems have different oracles $O_t$. For example, see \cite{jaquesImplementingGroverOracles2020} for the Advanced Encryption Standard decryption using Grover's algorithm. On the other hand, the implementation of diffusion operator $D_n$ is unambiguous. The $n$-qubit diffusion operator $D_n$ is equivalent to the $n$-qubit Toffoli gate (up to single-qubit gates) \cite{NC10}. The realization of $n$-qubit Toffoli gate on quantum computers is highly nontrivial. Quantum computers can only perform a set of basic single- and two-qubit gates, called the universal gate set. It is well-known that the $n$-qubit Toffoli gate can be decomposed into $\mathcal O(n)$ number of single- and two-qubit gates (requiring ancillary qubits) \cite{barencoElementaryGatesQuantum1995}. The specific number of the depth of $n$-qubit Toffoli gate depends on the connectivity and the type of universal gate set of quantum processors.

Grover firstly realized that the highly nonlocal diffusion operator can be replaced by the local diffusion operator \cite{groverTradeoffsQuantumSearch2002}. Here local diffusion operator means reflecting the part of amplitudes (divide the database into blocks). The local diffusion operator is a renormalized version of $D_n$, namely
\begin{equation}
    D_m = \left(1\!\!1_{2^m}-2|s_m\rangle\langle s_m|\right)\otimes 1\!\!1_{2^{n-m}},
\end{equation}
with integer $m\leq n$. The operator $D_m$ is equivalent to $m$-qubit Toffoli gate. If $m<n$, then realization of $D_m$ requires fewer gates (also fewer depths) than $D_n$. We can define the corresponding Grover operator as
\begin{equation}
    G_m = D_m O_t.
\end{equation}
To distinguish between $G_n$ and $G_m$, we call $G_n$ as the global Grover operator; $G_m$ as the local Grover operator. To be clarified, the local Grover operator still aims to solve the search problem with $N=2^n$. The oracle $O_t$ can not be renormalized for unstructured search problems. 

The simplest application of local Grover operator $G_m$ is to renormalize the database. Then we can prepare the state $|s_m\rangle\otimes |l\rangle$ as initial. Here the bit string $l\in \{0,1\}^{n-m}$ is randomly chosen. Correspondingly, the target state $|t\rangle$ is partitioned into $|t\rangle = |t_2\rangle\otimes |t_1\rangle$. Applying $G_m$ on $|s_m\rangle\otimes |l\rangle$ can find the target state $|t_2\rangle$ if $l=t_1$. Randomly choosing $l$ gives the probability $P_{l=t_1} = 1/2^{n-m}$. Then the total success probability of finding the target state is
\begin{equation}
P_{l=t_1}\tr\left(|t_2\rangle\langle t_2| G_m^j|s_m,l\rangle\langle s_m,l|{G_m^j}^\dag\right) = \frac{\sin^2((2j+1)\theta_b)}{2^{n-m}}
\end{equation}
with the short notation $|s_m,l\rangle = |s_m\rangle\otimes |l\rangle$. Here $\sin\theta_b = 1/\sqrt b$ and $b=2^m$. The algorithm combines the classical randomly guessing and the quantum search. Therefore, it is called the hybrid classical-quantum search algorithm \cite{zhangImplementationEfficientQuantum2021}. The optimal $j$ reaching the maximal success probability is much smaller than $\pi\sqrt N/4$. Therefore the circuit depth is dramatically reduced. The drawback is the success probability never approaches $1$. The circuit may need many rounds to find the target state. When $n-m$ is large, it is less efficient than the classical algorithm. 

A more sophisticated way to apply the local Grover operator is to replace some of $G_n$ with $G_m$ \cite{groverTradeoffsQuantumSearch2002, zhangDepthOptimizationQuantum2020a,brianskiIntroducingStructureExpedite2021,liuHardwareEfficientQuantum2021}. Since $G_n$ does not commute with $G_m$, the order of $G_n$ and $G_m$ is important. It provides flexibility to design quantum search algorithm which uses the least physical resource, such as depth. The nature of quantum algorithms is probabilistic. Therefore the proper metric characterizing the physical resource is the {\it expected depth} finding the target state. It is the average number of the depth needed to find the target state. There is always a trade-off between the circuit depth and width (number of qubits). For example, see \cite{kimTimeSpaceComplexity2018}. Currently, the circuit depth is a much more valuable resource than the circuit width. Therefore we do not consider the trade-off between circuit width (number of qubits) and depth here. 

Consider a sequence of operator
\begin{equation}
\label{eq:def_L_n_m}
    L_{n,m}(\gamma) = G_n^{j_1}G_m^{j_{2}}\cdots G_n^{j_{q-1}}G_m^{j_q},
\end{equation}
with $\gamma = \{j_1,j_2,\ldots,j_q\}$. Each local Grover operator $G_m$ acts on the same subspace of database. We denote the depth of $L_{n,m}(\gamma)$ as $d(L_{n,m}(\gamma))$. Different quantum processors (with different connectivities and universal gate sets) would have different $d(L_{n,m}(\gamma))$. The expected depth of finding the target state (using the operator $L_{n,m}(\gamma)$) is
\begin{equation}
\label{eq:def_L_expect}
    \langle L_{n,m}(\gamma) \rangle = \frac{d(L_{n,m}(\gamma))}{|\langle t|L_{n,m}(\gamma)|s_n\rangle|^2}.
\end{equation}
The optimal choice is to find the minimum of $\langle L_{n,m}(\gamma) \rangle$. The optimizations should go through all possible orders of $G_n$ and $G_m$. Besides, we can also choose different $m$ (the width of the local diffusion operator). Although replacing $D_n$ with $D_m$ will decrease the success probability, the saved depth (from $D_m$) may still reduce the expected depth $\langle L_{n,m}(\gamma) \rangle$. Theoretical study shows that Grover's algorithm is not optimal in depth as long as the depth of $D_n$ is not negligible compared to the depth of $O_t$. In Sec. \ref{subsec:benchmark_results}, we will show that implementations of $\langle L_{n,m}(\gamma) \rangle$ on NISQ devices not only reduce the depth, but also increase the success probability (compared to Grover's algorithm).

As mentioned in \ref{subsec:Grover}, Grover's algorithm becomes inefficient when $j$ approaches $\pi\sqrt N/4$. If we consider the expected depth of Grover's algorithm, $L_{n,m}(\gamma)$ with $m=n$ and $\gamma= \{j\}$, the minimum of expected depth is given by when $j$ approaches $0.5829\sqrt N$. Correspondingly, the minimal expected depth is around $0.6901\sqrt N$ times the depth of $G_n$. It is certainly smaller than $\pi\sqrt N/4$ times the depth of $G_n$.

\subsection{\label{subsec:divde_conquery}Divide-and-conquer quantum search algorithms}

Another application of local Grover operator $G_m$ is the quantum partial search algorithm \cite{groverPartialQuantumSearch2005,korepinSimpleAlgorithmPartial2006}. Quantum partial search algorithm trades accuracy for speed. It only finds partial bits of the target state, but the number of queries to oracle is smaller than Grover's algorithm. The original quantum partial search algorithm is designed by the operator $G_nG^{j_1}_mG^{j_2}_n$. The final measurement is performed on the qubits \textit{without} applying by the local diffusion operator. It finds $n-m$ bits of the target state almost with certainty. The iteration numbers $j_1$ and $j_2$ are optimized, by which gives the minimal number of queries to oracle (minimize the sum $j_1+j_2$).   

The quantum partial search algorithm naturally gives the framework of divide-and-conquer quantum search algorithms. At each stage, it finds partial bits of the target state. The last stage is a renormalized search problem, as discussed in Sec. \ref{subsec:depth_opt}. One may argue that the divide-and-conquer quantum search always increases the number of queries to the oracle, compared to Grover's algorithm. However, the divide-and-conquer strategy may provide an advantage for NISQ processors. In quantum search algorithms, circuits with longer depths imply higher success probabilities. However, the deep circuit may end in random results. Therefore it is more practical to design shorter depth circuits. It is also reasonable to expect the shorter depth circuit to find partial bits rather than the full target state. The divide-and-conquer quantum search algorithm may not be as efficient as Grover's algorithm, it still can provide a quantum advantage over the classical algorithm. 

The original quantum partial search algorithm is optimal in the number of queries to oracle \cite{korepinQuestFastPartial2006,korepinGroupTheoreticalFormulation2006}. However, it is not optimal in depth \cite{zhangDepthOptimizationQuantum2020a}. Similar as the ideas in Sec. \ref{subsec:depth_opt}, we can optimize the depth over through all possible configurations. Assume that we want to design a two-stage algorithm. The target state is partitioned into $|t\rangle = |t_2\rangle\otimes |t_1\rangle$. Consider the operator $L^{(1)}_{n,m}(\gamma_1)$, defined in Eq. (\ref{eq:def_L_n_m}), for the first stage. The probability that the operator $L^{(1)}_{n,m}(\gamma_1)$ finds the bit string $t_1$ is
\begin{equation}
    P^{(1)}_{n,m}(\gamma_1) = \tr\left(|t_1\rangle\langle t_1| L^{(1)}_{n,m}(\gamma_1) |s_n\rangle\langle s_n|{{L^{(1)}_{n,m}}^\dag(\gamma_1)}\right).
\end{equation}
Here the partial diffusion operator $D_m$ is acting on the first $m$ qubits. And we measure the last $n-m$ qubits. Note that the quantum partial search algorithm has the convention on measuring the qubits {\it without} acting by the partial diffusion operator. Only in special cases, measuring on qubits acted by the partial diffusion operator has a higher success probability. For more examples, see \cite{zhangDepthOptimizationQuantum2020a}. 

The second-stage circuit is a renormalized quantum search algorithm. Consider the operator $L^{(2)}_{n,k}(\gamma_2)$ with $k\leq m\leq n$. Assume that we find $t_1$ in the first stage. Then the probability finding $t_2$ is
\begin{multline}
    P^{(2)}_{n,k}(\gamma_2) = \\
\tr\left(|t_2\rangle\langle t_2|L^{(2)}_{n,k}(\gamma_2) |s_m,t_1\rangle\langle s_m,t_1|{{L^{(2)}_{n,k}}^\dag(\gamma_2)}\right),
\end{multline}
with the short notation $|s_m,t_1\rangle = |s_m\rangle\otimes |t_1\rangle$. Then we can find the expected depth of above two-stage search algorithm, given by
\begin{equation}
\label{eq:def_L_expect_two_stage}
    \langle L^{(1)}_{n,m}(\gamma_1) + L^{(2)}_{n,k}(\gamma_2)\rangle = \frac{d\left(L^{(1)}_{n,m}(\gamma_1)\right)+d\left(L^{(2)}_{n,k}(\gamma_2)\right)}{P^{(1)}_{n,m}(\gamma_1)P^{(2)}_{n,k}(\gamma_2)}.
\end{equation}
The optimal realization is to choose $m,k,\gamma_1,\gamma_2$ which gives the minimum $\langle L^{(1)}_{n,m}(\gamma_1) + L^{(2)}_{n,k}(\gamma_2)\rangle$. If the oracle can not verify the partial bits, we can not separately optimize the expected depth of $L^{(1)}_{n,m}(\gamma_1)$ or $L^{(2)}_{n,k}(\gamma_2)$.

Surprisingly, when $d(O_t)<(1+\sqrt 3)d(D_n)$, the minimal expected depth of two-stage algorithm can still be smaller than Grover's algorithm (assuming no errors in quantum processors) \cite{zhangDepthOptimizationQuantum2020a}. The previous study has shown the advantage of two-stage four-qubit quantum search algorithm on IBMQ devices \cite{zhangImplementationEfficientQuantum2021}.

\section{\label{sec:benchmark_results} Quantum-search-oriented Benchmarks}

In this section, we present the results of the five-qubit quantum search algorithms on IBMQ, IonQ and Honeywell quantum devices. The circuits include the original Grover's algorithm, the depth-optimized, and the divide-and-conquer quantum search algorithms. First, we introduce the benchmark metrics in Sec. \ref{subsec:benchmark_metrics}. We present the search circuits in Sec. \ref{subsec:benchmark_circuits}. The benchmark results are shown in Sec. \ref{subsec:benchmark_results}. In Sec. \ref{subsec:benchmark_sim}, we compare the results from real quantum devices with the noisy simulations. 

\subsection{\label{subsec:benchmark_metrics}Benchmark metrics}

The nature of quantum algorithms is probabilistic. The output probabilities are estimated by running the circuit multiple shots. To benchmark the performance of quantum search algorithms, we choose the following metrics. 
\begin{itemize}
    \item Success probability. The success probability of finding the target state is the most important metric for quantum search algorithms. 
    
    \item Expected depth. It is defined as the depth over the success probability, such as shown in Eqs. (\ref{eq:def_L_expect}) and (\ref{eq:def_L_expect_two_stage}). It characterizes the average depth for finding the target state. 
    
    \item Selectivity. Selectivity was introduced in \cite{tannuEnsembleDiverseMappings2019,wangProspectUsingGrover2020}. It is defined as
    \begin{equation}
        S = \ln\left(\frac{P_t}{\max P_{nt}}\right),
    \end{equation}
    assuming $\max P_{nt}\neq 0$. Here $P_t$ is the probability finding the target state; $P_{nt}$ is the probability finding the non-target state. The logarithmic function is added for convenience. Since the amplitude of the non-target state is never amplified, a negative selectivity in quantum search algorithms means that it is worse than the random guess. Higher selectivity implies that the circuit is more robust to noise (to prevent becoming random noises). 
    
    There is one special case, the two-qubit search (or the rescaled two-qubit search). It is the only case where Grover's algorithm finds the target state with success probability $1$ \cite{diaoExactnessOriginalGrover2010}. Then we define the corresponding selectivity as infinity. 
    
    \item Circuit fidelity. The circuit fidelity is a universal benchmark metric introduced in \cite{lubinskiApplicationOrientedPerformanceBenchmarks2021}. It quantifies the distance between the output probabilistic distribution $P_\text{out}$ with the ideal $P_\text{ideal}$. Besides, it is normalized in terms of the uniform distribution $P_\text{uni}$. The definition is 
    \begin{multline}
    \label{eq:def_fidelity}
        \hspace{0.6cm} F(P_\text{out},P_\text{ideal}) = \\ \frac{f(P_\text{out},P_\text{ideal})-f(P_\text{uni},P_\text{ideal})}{1-f(P_\text{uni},P_\text{ideal})},
    \end{multline}
    with the distance
    \begin{equation}
        f(P_1,P_2) = \left(\sum_j \sqrt{P_1(j)P_2(j)}\right)^2.
    \end{equation}
    Obviously, we have $F(P_\text{out},P_\text{ideal})\leq 1$ and $F(P_\text{ideal},P_\text{ideal})=1$. If the output is uniform, then $F(P_\text{uni},P_\text{ideal}) = 0$. Not that $F(P_\text{out},P_\text{ideal})$ could also be negative. It reflects that the output is more distant to the ideal distribution than the uniform distribution. When the noises are dominated, the output might not be uniformed. For example, the qubits will relax to a nonuniform distribution due to the thermal decoherence \cite{NC10}.
    
    Note that the high fidelity does not necessarily mean the high success probability. The circuit fidelity quantifies the degree of degradation due to the noises. As argued in \cite{lubinskiApplicationOrientedPerformanceBenchmarks2021}, the metric success probability and circuit fidelity are complementary for benchmarking.

\end{itemize}

\subsection{\label{subsec:benchmark_circuits}Benchmark circuits}

Previous studies on the implementation of Grover's algorithm focus on the four-qubit search domain \cite{mandviwallaImplementingGroverAlgorithm2018,hlembotskyiEfficientUnstructuredSearch2020,satohSubdividedPhaseOracle2020,gwinnerBenchmarking16elementQuantum2021,zhangImplementationEfficientQuantum2021}. In our study, we march onto the five-qubit quantum search. Not only the quantum processors are improving, the depth optimization and divide-and-conquer strategies also improve the performance of quantum search algorithms. We design the following circuits.
\begin{itemize}
    \item G5M5: apply one $G_5$ operator, and then measure all the five qubits. Here G and M stand for the Grover operator and the measurement respectively. Like the convention of quantum circuit diagrams, the notation G5M5 reads from left to right. G5M5 is the one-iteration five-qubit Grover's algorithm. The corresponding quantum circuit diagram is
    $$
    \Qcircuit @C=1em @R=1em {
    & |0\rangle && \gate{H} & \multigate{4}{O_t} & \multigate{4}{D_5} & \meter \\
    & |0\rangle && \gate{H} & \ghost{O_t} & \ghost{D_5} & \meter \\
    & |0\rangle && \gate{H} & \ghost{O_t} & \ghost{D_5} & \meter \\
    & |0\rangle && \gate{H} & \ghost{O_t} & \ghost{D_5} & \meter \\
    & |0\rangle && \gate{H} & \ghost{O_t} & \ghost{D_5} & \meter 
    }
    $$
    
    \item G5G5M5: apply two $G_5$ operators, and then measure all the five qubits. G5G5M5 is the two-iteration five-qubit Grover's algorithm. The corresponding circuit diagram is
    $$
    \Qcircuit @C=1em @R=1em {
    & |0\rangle && \gate{H} & \multigate{4}{O_t} & \multigate{4}{D_5} & \multigate{4}{O_t} & \multigate{4}{D_5} & \meter \\
    & |0\rangle && \gate{H} & \ghost{O_t} & \ghost{D_5} & \ghost{O_t} & \ghost{D_5} & \meter \\
    & |0\rangle && \gate{H} & \ghost{O_t} & \ghost{D_5} & \ghost{O_t} & \ghost{D_5} & \meter \\
    & |0\rangle && \gate{H} & \ghost{O_t} & \ghost{D_5} & \ghost{O_t} & \ghost{D_5} & \meter \\
    & |0\rangle && \gate{H} & \ghost{O_t} & \ghost{D_5} & \ghost{O_t} & \ghost{D_5} & \meter 
    }
    $$
    
    \item R2G3M3: randomly initialize the values of two qubits then apply the $G_3$ operator and measure the rest three qubits. The corresponding circuit diagram is 
    $$
    \Qcircuit @C=1em @R=1em {
    & |0\rangle && \gate{X^{l_1}} & \multigate{4}{O_t} & \qw \\
    & |0\rangle && \gate{X^{l_2}} & \ghost{O_t} & \qw \\
    & |0\rangle && \gate{H} & \ghost{O_t} & \multigate{2}{D_3} & \meter \\
    & |0\rangle && \gate{H} & \ghost{O_t} & \ghost{D_3} & \meter \\
    & |0\rangle && \gate{H} & \ghost{O_t} & \ghost{D_3} & \meter 
    }
    $$
    Here $X$ is the Pauli-$x$ gate and $l_1,l_2\in\{0,1\}$ is randomly chosen. 
    
    \item R3G2M2: randomly initialize the values of three qubits then apply the $G_2$ operator and measure the rest two qubits. The corresponding circuit diagram is 
    $$
    \Qcircuit @C=1em @R=1em {
    & |0\rangle && \gate{X^{l_1}} & \multigate{4}{O_t} & \qw \\
    & |0\rangle && \gate{X^{l_2}} & \ghost{O_t} & \qw  \\
    & |0\rangle && \gate{X^{l_3}} & \ghost{O_t} & \qw  \\
    & |0\rangle && \gate{H} & \ghost{O_t} & \multigate{1}{D_2} & \meter \\
    & |0\rangle && \gate{H} & \ghost{O_t} & \ghost{D_2} & \meter 
    }
    $$
    
    \item G2M2$\mid$G3M3: in the first stage, apply the $G_2$ operator then measure two qubits; in the second stage, initialize the two qubits based on the first-stage measurement results, then apply the $G_3$ operator and measure the rest three qubits. We adopt the notation ``$\mid$'' to separate two stages. We can recycle the qubits after implementing the first-stage circuit. Note that the second stage circuit is equivalent to R2G3M3. The corresponding circuit is
    
    $$
    \Qcircuit @C=0.6em @R=0.6em {
    & |0\rangle && \gate{H} & \multigate{4}{O_t} & \qw  \\
    & |0\rangle && \gate{H} & \ghost{O_t} & \qw  \\
    & |0\rangle && \gate{H} & \ghost{O_t} & \qw  \\
    & |0\rangle && \gate{H} & \ghost{O_t} & \multigate{1}{D_2} & \qw & \meter \\
    & |0\rangle && \gate{H} & \ghost{O_t} & \ghost{D_2} & \meter \\
    &&&&& |0\rangle~~~~~~ & \gate{X} \cwx & \qw     & \multigate{4}{O_t} & \qw \\
    &&&&& |0\rangle~~~~~~ & \qw & \gate{X} \cwx[-3] & \ghost{O_t} & \qw \\ 
    &&&&& |0\rangle~~~~~~ & \qw & \gate{H} & \ghost{O_t} & \multigate{2}{D_3} & \meter \\ 
    &&&&& |0\rangle~~~~~~ & \qw & \gate{H} & \ghost{O_t} & \ghost{D_3} & \meter \\ 
    &&&&& |0\rangle~~~~~~ & \qw & \gate{H} & \ghost{O_t} & \ghost{D_3} & \meter \\ 
    }
    $$
    
    \item G3M3$\mid$G2M2: in the first stage, apply the $G_3$ operator then measure three qubits; in the second stage, initialize the three qubits based on the first-stage measurement results, then apply the $G_2$ operator and measure the rest two qubits. The second stage circuit is equivalent to R3G2M2. The corresponding circuit diagram is
    
    $$
    \Qcircuit @C=0.6em @R=0.6em {
    & |0\rangle && \gate{H} & \multigate{4}{O_t} & \qw  \\
    & |0\rangle && \gate{H} & \ghost{O_t} & \qw  \\
    & |0\rangle && \gate{H} & \ghost{O_t} & \multigate{2}{D_3} & \qw & \qw & \meter  \\
    & |0\rangle && \gate{H} & \ghost{O_t} & \ghost{D_3} & \qw & \meter \\
    & |0\rangle && \gate{H} & \ghost{O_t} & \ghost{D_3} & \meter \\
    &&&&& |0\rangle~~~~~~ & \gate{X} \cwx & \qw     & \qw & \multigate{4}{O_t} & \qw \\
    &&&&& |0\rangle~~~~~~ & \qw & \gate{X} \cwx[-3] & \qw & \ghost{O_t} & \qw \\ 
    &&&&& |0\rangle~~~~~~ & \qw & \qw & \gate{X} \cwx[-5] & \ghost{O_t} & \qw \\ 
    &&&&& |0\rangle~~~~~~ & \qw & \qw & \gate{H} & \ghost{O_t} & \multigate{1}{D_2} & \meter \\ 
    &&&&& |0\rangle~~~~~~ & \qw & \qw & \gate{H} & \ghost{O_t} & \ghost{D_2} & \meter \\ 
    }
    $$
    
\end{itemize}

\quad

Different search problems have different oracles. As a toy oracle, we set the oracle as a five-qubit controlled phase gate. It is also single-qubit-gate equivalent to the five-qubit Toffoli gate. See Appendix \ref{App:circuits} for its implementation (with the help of the relative-phase Toffoli gate \cite{maslovAdvantagesUsingRelativephase2016}). Quantum processors with different connectivity will have different implementations of the five-qubit Toffoli gate. Physical qubits tend to relax to the ground state $|0\rangle$, then target states with many zeros will have higher success probabilities. To exclude any bias related to the target state, we randomly choose the target state as $|01011\rangle$. For more detailed setups on the IBMQ, IonQ, and Honeywell quantum devices, see Appendix \ref{App:setup}.

\subsection{\label{subsec:benchmark_results}Benchmark results}

We implement the quantum search circuits designed in the last section on four different quantum processors: IBMQ Lagos, IBMQ Mumbai, IonQ, and Honeywell System Model H1. IBMQ Lagos and IBMQ Mumbai have seven and twenty-seven superconducting qubits respectively. IonQ and Quantinuum System Model H1 are both trapped-ion quantum processors. IonQ has eleven fully connected physical qubits. Quantinuum System Model H1 has twelve qubits, also all fully connected. 

The common metric benchmarking quantum processors is the quantum volume \cite{crossValidatingQuantumComputers2019}. Quantum volume equals to $2^k$, where $k$ is the largest number that $k$-width and $k$-depth random circuits can be successfully executed (the average fidelity is larger than a threshold). IBMQ Lagos and IBMQ Mumbai have the quantum volume $32$ and $128$ respectively \cite{IBM}. Honeywell System Model H1 has reported the quantum volume $1024$ \cite{Quantinuum}. In \cite{lubinskiApplicationOrientedPerformanceBenchmarks2021}, the estimated quantum volume of IonQ system is $64$.

\begin{figure}[t!]
\includegraphics[width=\columnwidth]{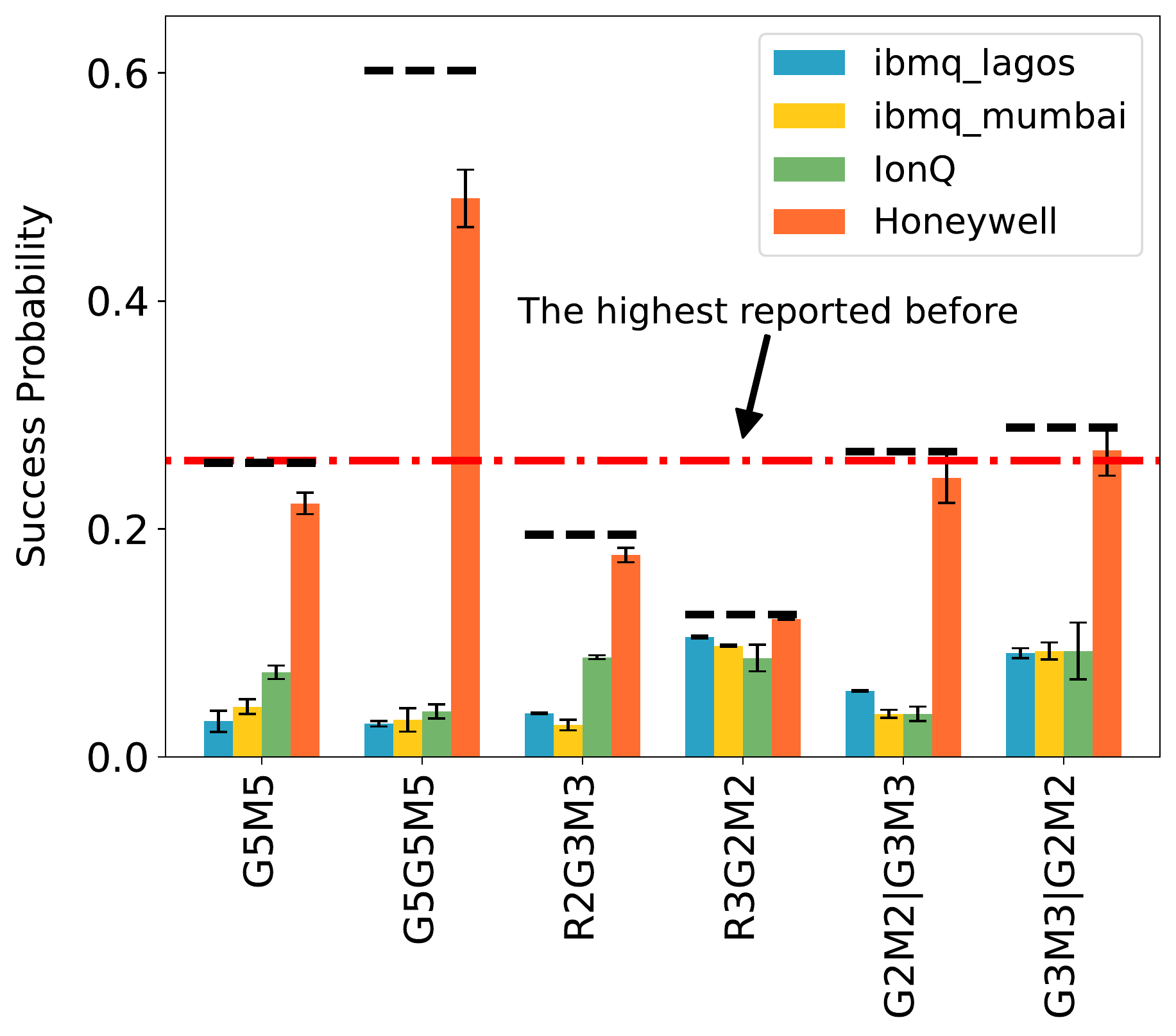}
\caption{Success probabilities of the quantum search circuits G5M5, G5G5M5, R2G3M3, R3G2M2, G2M2$\mid$G3M3, and G3M3$\mid$G2M2. The black dashed lines represent the theoretical success probabilities. The red dash-dotted line represents the highest success probability of the five-qubit search reported in \cite{hlembotskyiEfficientUnstructuredSearch2020}. The variance is based on $3\times 400$ shots.}
\label{fig_suc_prob}
\end{figure}

The success probabilities of all circuits are shown in Fig. \ref{fig_suc_prob}. Circuits G5M5, R2G3M3, and R3G2M2 only use one oracle. G5G5M5, G2M2$\mid$G3M3, and G3M3$\mid$G2M2 are two-oracle circuits. Among the six quantum search circuits, the circuit R3G2M2 gives the maximal success probability on both IBMQ Lagos and Mumbai. Even though IBMQ Mumbai has a larger quantum volume than Lagos, it does not provide any advantage on the five-qubit quantum search algorithms. For IonQ, the maximum is given by the two-stage circuit G3M3$\mid$G2M2. Honeywell quantum device gives the success probability around 0.49 for the two-iteration Grover's algorithm G5G5M5. It is the highest among all the implementations. As far as the authors' knowledge, it is also higher than the success probabilities reported before \cite{hlembotskyiEfficientUnstructuredSearch2020,zhangImplementationEfficientQuantum2021}. See Fig. \ref{fig_suc_prob} for the comparison. Due to the shallow depth of R3G2M2, its outputs are always very close to the theoretical values. Note that circuit R3G2M2 is still superior to the classical search with one query to oracle. 

\begin{figure}[t!]
\includegraphics[width=\columnwidth]{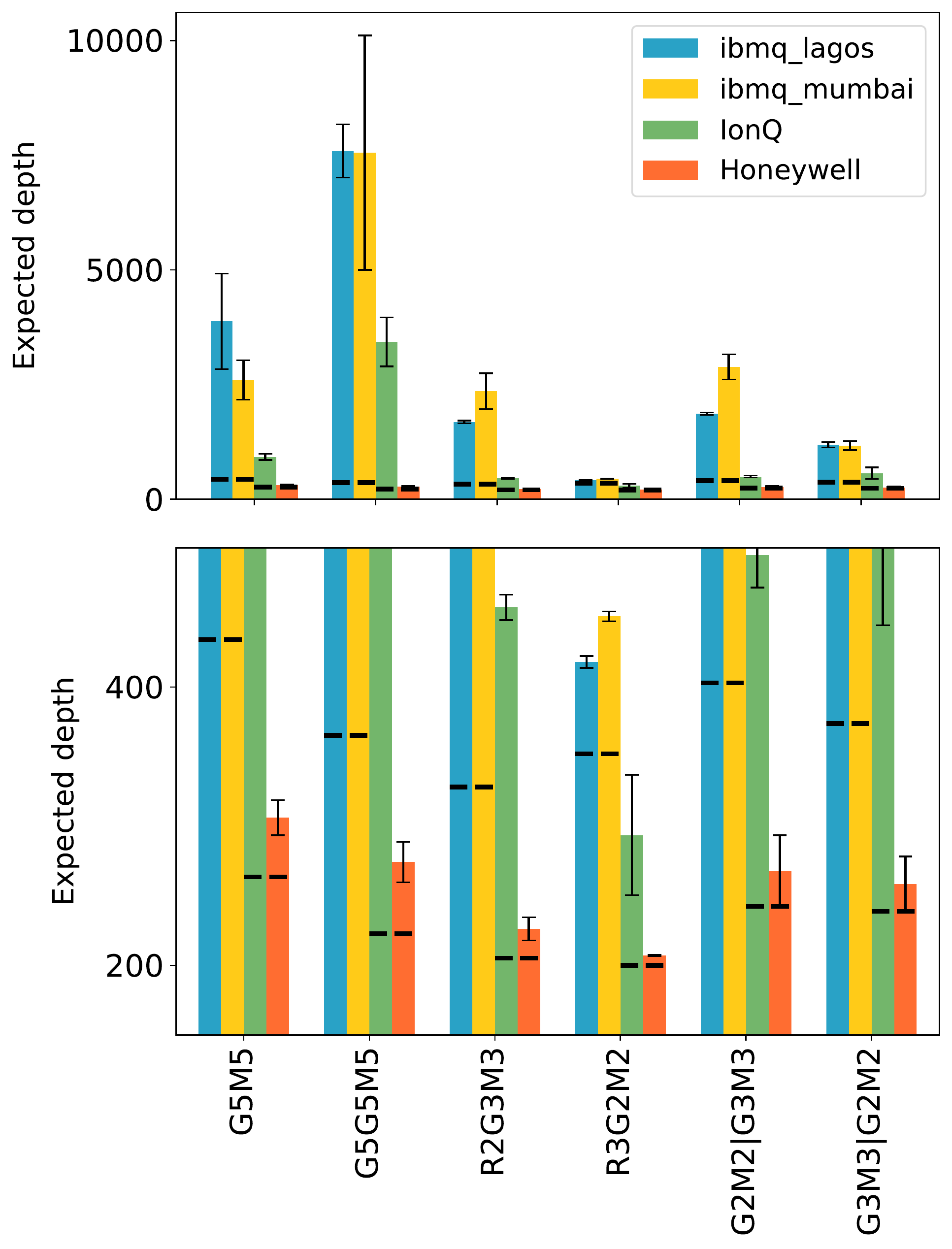}
\caption{Expected depths of the quantum search circuits G5M5, G5G5M5, R2G3M3, R3G2M2, G2M2$\mid$G3M3, and G3M3$\mid$G2M2. The above figure is the full view of the plot. The bottom figure is plotted in the range of 150 to 500 depths. The dashed lines represent the theoretical values. The variance is based on $3\times 400$ shots.}
\label{fig_depth}
\end{figure}

IBM Quantum processors Lagos and Mumbai have the same universal gate set. Although IonQ and Honeywell quantum processors are both based on the trapped-ion qubits, their universal gate sets are different. Qiskit (IBMQ programming software) can directly report the circuit depth (after compilation), which is the accurate depth implemented on the hardware. The circuit depths on Lagos and Mumbai are the same since we choose the physical qubits which have the same connectivity on those two devices. More details can be found in Appendix \ref{App:setup}. We estimate the circuit depths implemented on IonQ and Honeywell quantum devices via the universal gate set including CNOT and arbitrary single-qubit gates. The native two-qubit gates provided by IonQ and Honeywell are both single-qubit-gate equivalent to CNOT. Therefore the estimated depth is only slightly shorter than the actual depth. Since both the qubits of IonQ and Honeywell are fully connected. The same circuits have the same estimated depth on those two devices. We report the following depth for the six search circuits: G5M5 (112, 68), G5G5M5 (220,134), R2G3M3 (64,40), R3G2M2 (44,25), G2M2$\mid$G3M3 (108,65) and G3M3$\mid$G2M2 (108,69). Here the first number is the depth for IBMQ Lagos and Mumbai; the second number is for IonQ and Honeywell. The depth of R3G2M2 is only around one-third of the one-iteration Grover's algorithm. 

\begin{figure}[t!]
\includegraphics[width=\columnwidth]{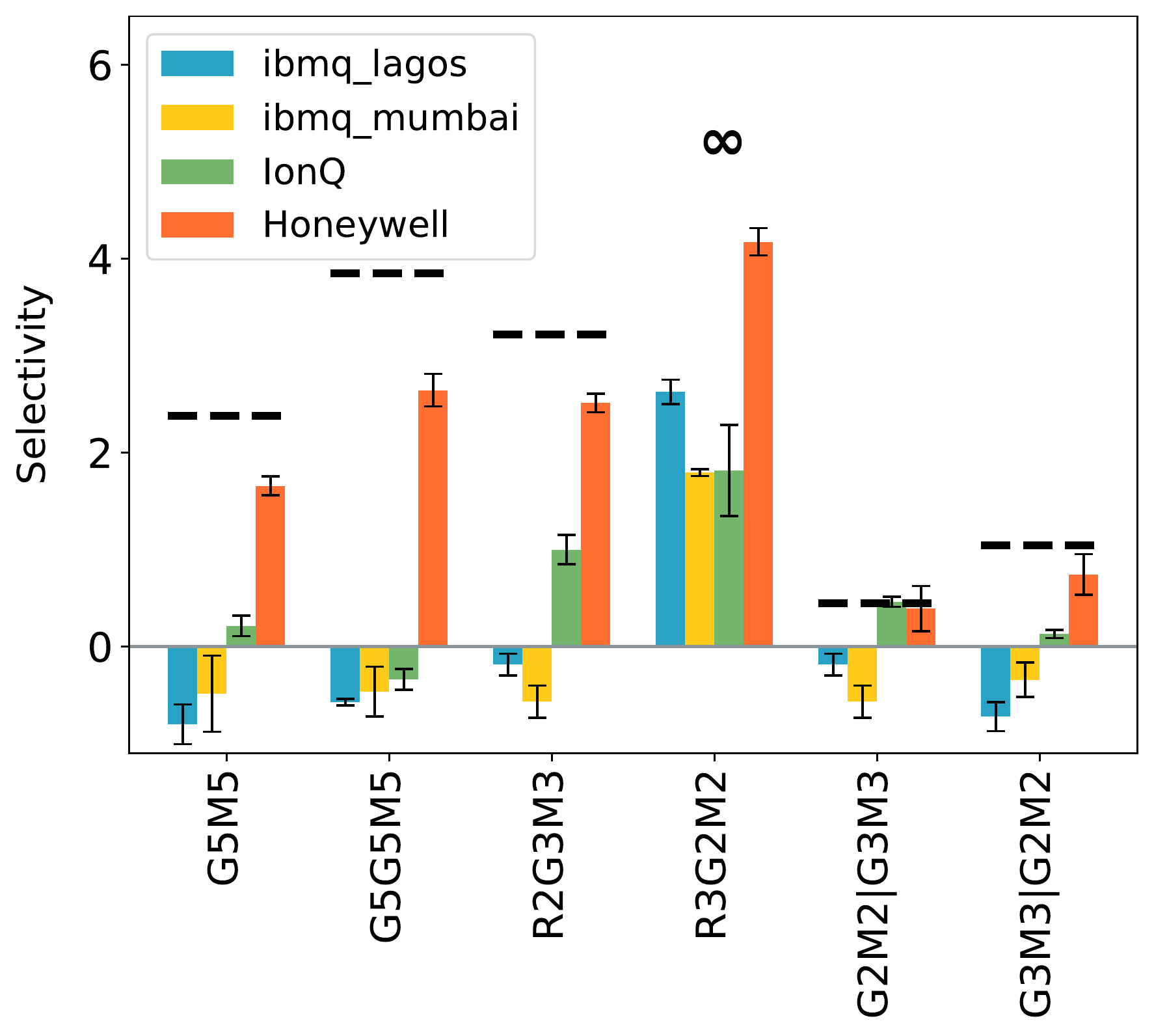}
\caption{Selectivities of the quantum search circuits G5M5, G5G5M5, R2G3M3, R3G2M2, G2M2$\mid$G3M3, and G3M3$\mid$G2M2. The dashed lines represent the theoretical values. Circuit R3G2M2 has the infinite selectivity. The variance is based on $3\times 400$ shots.}
\label{fig_sele}
\end{figure}

We report the expected depth in Fig. \ref{fig_depth}. The theoretically expected depth for IBMQ Lagos and Mumbai is larger than the ones for IonQ and Honeywell, because of the extra SWAP gates required for IBMQ Lagos and Mumbai. From Fig. \ref{fig_depth}, we can obviously see that R3G2M2 has the lowest expected depth (both in theory and practice). Moreover, the expected depth of R3G2M2 is smaller than the one- and two-iteration Grover's algorithm. It is similar as reported in \cite{zhangImplementationEfficientQuantum2021} for four-qubit search algorithm. As analyzed in theory \cite{zhangDepthOptimizationQuantum2020a}, Grover's algorithm is not optimal in depth. Therefore it is practical to design quantum search circuits which trade the success probability for circuit depths, either for NISQ or post-NISQ devices.

Selectivity reflects the ability to infer the correct target state from the output distribution \cite{tannuEnsembleDiverseMappings2019,wangProspectUsingGrover2020}. There would be two selectivities in the two-stage circuit. We choose the smaller one as the selectivity for the two-stage circuit. The results on selectivity are shown in Fig. \ref{fig_sele}. Negative selectivity means that there is a non-target state having a higher probability than the target state. It means the failure of the algorithm. The theoretical selectivity of R3G2M2 is infinitely large since it is a renormalized two-qubit search circuit (with success probability one). The five-qubit search algorithm is the border for IBMQ devices. The IonQ device can afford one-iteration Grover's algorithm. The Honeywell quantum device can successfully find the target state in all six different search circuits. 

\begin{figure}[t!]
\includegraphics[width=\columnwidth]{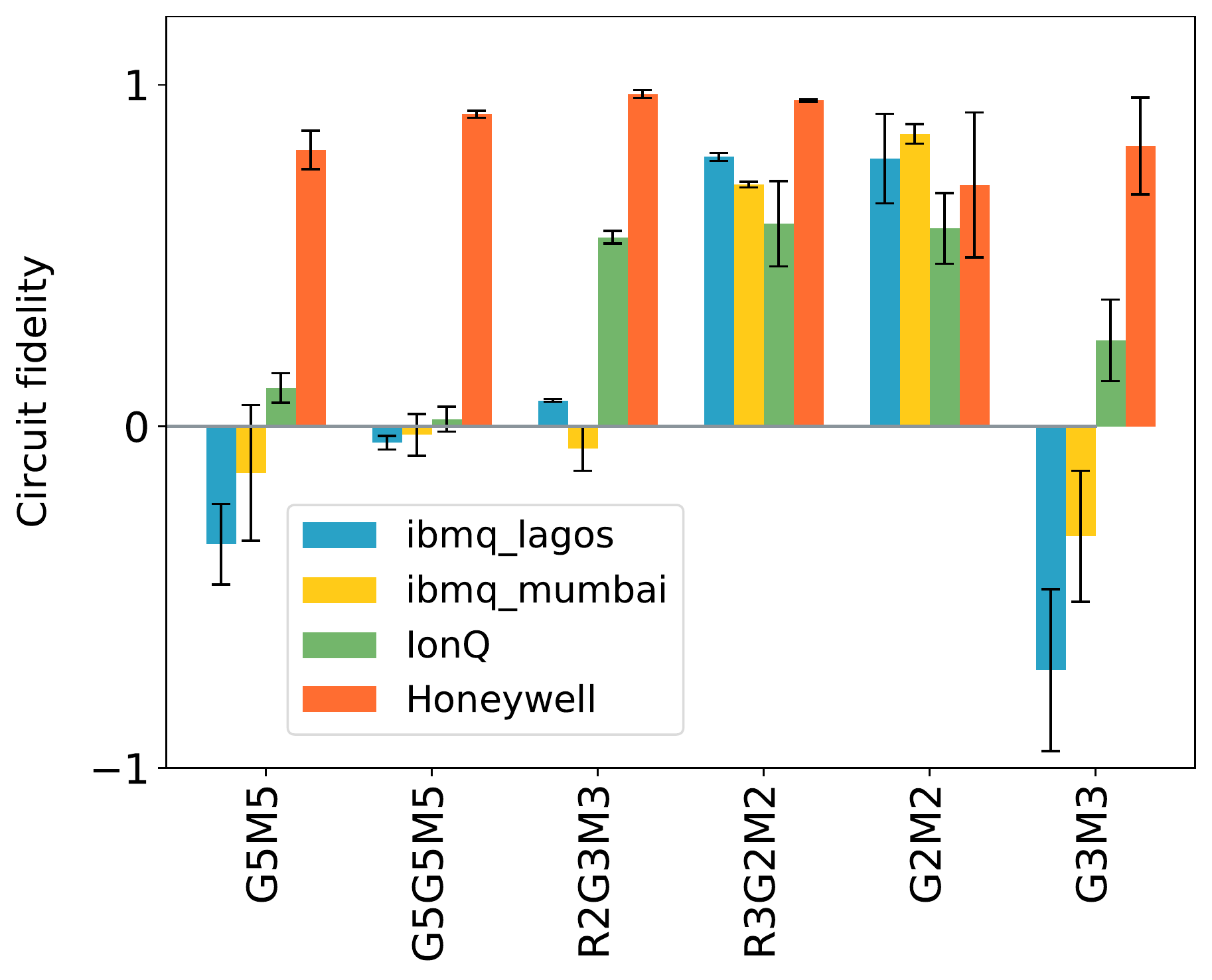}
\caption{Circuit fidelities of the quantum search circuits G5M5, G5G5M5, R2G3M3, R3G2M2, G2M2$\mid$G3M3 and G3M3$\mid$G2M2. The variance is based on $3\times 400$ shots.}
\label{fig_fidelity}
\end{figure}

Circuit fidelity defined in Eq. (\ref{eq:def_fidelity}) characterizes the distance between the output and ideal distributions. Recall that the perfect match gives the fidelity one; the uniform distribution gives the fidelity zero. Since the second stage circuit of G2M2$\mid$G3M3 is equivalent to R2G3M3, here we only benchmark the fidelity of the first stage circuit G2M2. Similar rules apply to G3M3$\mid$G2M2. For Honeywell quantum devices, relative deep circuits, such as G5G5M5, does not decrease much of fidelity compared to other circuits. Honeywell quantum devices can handle well on circuits with over two hundred depths (around a hundred depths contain two-qubit gates). Circuits R3G2M2 and G2M2 have only one two-qubit diffusion operator. All processors can give fidelity over 0.5 on those two circuits. Circuits with negative fidelities are consistent with the negative selectivities shown in Fig. \ref{fig_sele}.

\subsection{\label{subsec:benchmark_sim}Benchmark by noisy simulations}

\begin{figure*}[t!]
\includegraphics[width=\textwidth]{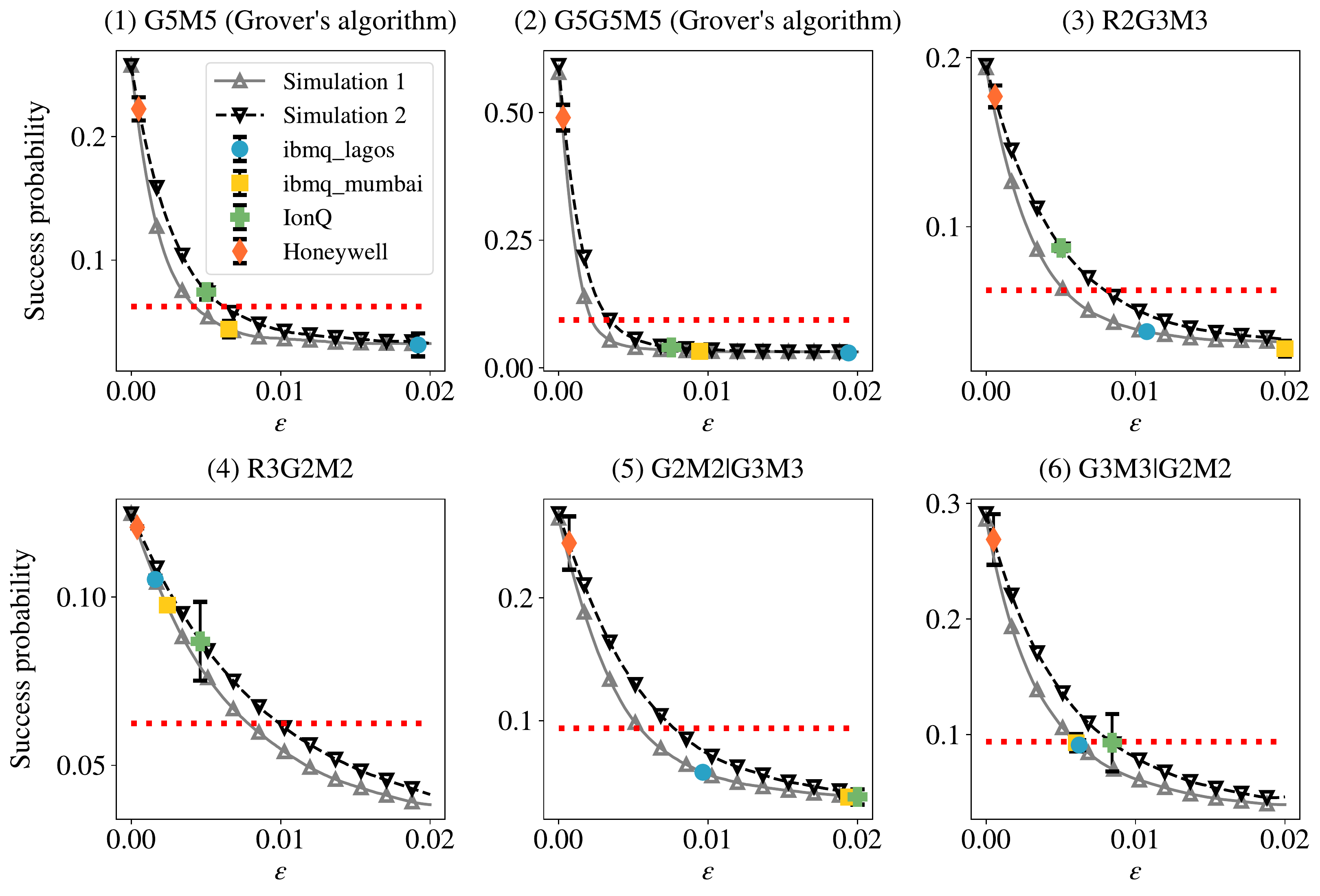}
\caption{Benchmark on the quantum search circuits G5M5, G5G5M5, R2G3M3, R3G2M2, G2M2$\mid$G3M3, and G3M3$\mid$G2M2 by noisy simulations. Simulation 1 is based on the limited qubits connectivity (based on the qubits layout of IBMQ Lagos and Mumbai). Simulation 2 is based on the fully connected qubits. The red dotted line is the classical success probability of finding the target state (given by one or two queries to the oracle). }
\label{fig_nosiy_sim}
\end{figure*}

To have a more concrete understanding of the benchmark results presented in the last section, we compare the measured results with the noisy simulations. An accurate noisy simulation may need dozen of parameters to characterize different types of errors, such as the noisy model in \cite{georgopoulosModellingSimulatingNoisy2021}. On the other hand, we can compare the results from the real devices with a simple error model, in which the strength of noises is a free parameter. In this way, the behavior of noisy simulation is more tractable. The most common error model is the depolarizing channel, denoted as $\mathcal L(\rho)$ \cite{NC10}. It has the form
\begin{equation}
    \mathcal L(\rho) = \left(1-\varepsilon\right)\rho + \frac{\varepsilon}{2^n} 1\!\!1, 
\end{equation}
with the error rate $\varepsilon\in [0,1]$ and the number of qubits denoted as $n$. The depolarizing channel characterizes the decoherece of the qubits. The steady-state of depolarizing channel is the completely mixed state. It predicts the uniform distribution of any deep circuit. 

In the noisy simulation, we impose the depolarizing channel after each gate. We set the error rate $\varepsilon$ of two-qubit gates as ten times than the error rate of single-qubit gates. The ten times ratio is the typical error rate reported in NISQ devices. And it is also applied in \cite{lubinskiApplicationOrientedPerformanceBenchmarks2021}. Based on the above setups, we generate the noisy simulation via the Qiskit simulator. Here the error rate $\varepsilon$ is a variable. By varying the error rate, we export the success probability for each quantum search circuit. Then fit the measured results on the simulation curves. Therefore we can benchmark the error rates based on the measured success probability. The results are shown in Fig. \ref{fig_nosiy_sim}. There are two simulation curves. Simulation 1 (the bottom dashed curve) assumes the full connectivity of qubits. Simulation 2 (the below solid curve) is based on the limited connectivity of qubits. And the limited connectivity is based on the processors of IBMQ Lagos and Mumbai. For comparison, the classical success probability (with one or two queries to oracle) is also marked in Fig. \ref{fig_nosiy_sim} (red dotted line).

First, it is expected that the simulation with full connected qubits has a higher success probability than the limited connectivity case. Second, simulations on all the six search circuits show that the error rate of Honeywell quantum devices is below $0.1\%$. Correspondingly, the error rate of the two-qubit gate is below $1\%$. It is consistent with the gate fidelity reported by Honeywell. Third, we can also read the threshold error rate, below which gives the success probability lower than the classical case. All the four modified search circuits (using the partial diffusion operator) have higher threshold error rates compared to Grover's algorithm. Especially, the threshold error rate of circuit R3G2M2 is two times as high as the threshold error rate of one-iteration Grover's algorithm G5M5. Because of such an advantage of the circuit R3G2M2, it is the only circuit that all four processors report a higher success probability than the classical value. See Fig. \ref{fig_nosiy_sim} (4). 

\section{\label{sec:conclusion}Conclusions and Outlooks}

In this study, we present the benchmark of five-qubit quantum search algorithms on different NISQ devices, including quantum processors from IBMQ, IonQ, and Honeywell. Besides Grover's algorithm, we include the depth-optimized and divide-and-conquer search circuits in our study \cite{zhangDepthOptimizationQuantum2020a}. Our five-qubit search algorithms achieve the highest success probability on all three (IBMQ, IonQ, and Honeywell) quantum processors, compared to the previous works \cite{hlembotskyiEfficientUnstructuredSearch2020,zhangImplementationEfficientQuantum2021}. Honeywell quantum processors can successfully run the five-qubit Grover's algorithm as well as the quantum search circuits designed for NISQ devices. However, IBMQ and IonQ processors can not afford the long depth of the five-qubit Grover's algorithm. Therefore, depth-optimized quantum circuits are more practical. We also show that the depth of Grover's algorithm is not optimal, both in theory and experiments. Replacing the global diffusion operator by the local diffusion operator may decrease the actual success probability (not necessarily though). But it saves the physical resource for finding the target state. 

To understand and extract the full power of near-term quantum computers, designing more practical implementations is necessary. There is still a long way to run classical-impossible search tasks on quantum computers, such as the AES decryption. But mid-size quantum search algorithm, such as $N=2^{30}$, may serve as a subroutine, to provide a quantum speedup in near future. How to optimize the quantum search algorithms (finding the minimal expected depth or the minimal $T$-depth \cite{amyMeetinthemiddleAlgorithmFast2013}) running on fifty qubits is a highly nontrivial problem, since the circuits may not be classically tractable. It is also an open question on how to determine the divide-and-conquer strategy for the quantum search algorithms. Note that divide-and-conquer strategy also means that we can parallel run the algorithm \cite{gingrichGeneralizedQuantumSearch2000}. It is also interesting to explore how the local diffusion operator influences the coherence and the success probability \cite{pan2022complementarity}. We leave those questions for future study.

\begin{acknowledgments}
The authors thank Tzu-Chieh Wei, Yusheng Zhao, and Hongye Yu for helpful discussions, especially for Yusheng Zhao who pinpointed the bug in using Qiskit for IonQ devices. This research used quantum computing resources of the Oak Ridge Leadership Computing Facility, which is a DOE Office of Science User Facility supported under Contract DE-AC05-00OR22725. This research used resources of the National Energy Research Scientific Computing Center, a DOE Office of Science User Facility supported by the Office of Science of the U.S. Department of Energy under Contract No. DE-AC02-05CH11231 using NERSC award DDR-ERCAP0022229.
We acknowledge the use of an IonQ quantum computer, provided through the IonQ Research Credit Program and PROVIDER, for this work.  We acknowledge the access to IonQ and Honeywell Quantum Solution through the Microsoft Azure Quantum grant No. 35984.
	
\end{acknowledgments}

\appendix

\section{\label{App:circuits} Implementations of five-qubit Toffoli gate}

In our study, we choose the five-qubit controlled phase gate (up to some single-qubit gates) as the oracle. It is equivalent to the five-qubit generalized Toffoli gate up to single-qubit gates. Besides, the five-qubit diffusion operator is also single-qubit-gate-equivalent to the five-qubit generalized Toffoli gate. Although decomposing the generalized Toffoli gate without any ancillary qubits is possible, the decomposition will have shorter depth if clean ancillary qubits are permitted \cite{NC10}. In the following, we denote the $n$-qubit controlled gate as $\Lambda_{n-1}(U)$ (as there are $n-1$ controlled qubits), where $U$ is the controlled operation. For example, the five-qubit Toffoli gate is $\Lambda_4(X)$. 

For $\Lambda_4(X)$, we design the circuit 
$$
\Qcircuit @C=0.7em @R=1em {
& \ctrl{1} & \qw &&&&&&& \ctrl{1} & \qw & \ctrl{1} & \qw \\
& \ctrl{1} & \qw &&&&&&& \ctrl{1} & \qw & \ctrl{1} & \qw \\
& \ctrl{2} & \qw &&&&&&& \ctrl{1} & \qw & \ctrl{1} & \qw \\
& \qw \qw& \qw &&& \raisebox{0.5cm}{=} \hspace{0.5cm} & \hspace{0.4cm} |0\rangle &&& \gate{Y} & \ctrl{1} & \gate{Y^\dag} & \qw \\
& \ctrl{1} & \qw &&&&&&& \qw & \ctrl{1} & \qw & \qw \\
& \targ \qw & \qw &&&&&&&  \qw & \targ \qw & \qw & \qw   
}
\vspace{-0.2cm}
$$
Here $Y = |0\rangle\langle1|-|1\rangle\langle 0|$.  In the above implementations, the gates $\Lambda_3(Y)$ and $\Lambda_3(Y^\dag)$ can be replaced by the Toffoli gate $\Lambda_3(X)$. They only differ a relative sign. However, $\Lambda_3(Y)$ is more easily realized (by fewer depths) than $\Lambda_3(X)$ \cite{maslovAdvantagesUsingRelativephase2016}. We can decompose $\Lambda_3(Y)$ as 
\begin{widetext}
$$
\Qcircuit @C=0.8em @R=1.2em {
& \ctrl{1} & \qw &&&& \qw & \qw & \qw & \qw & \qw & \ctrl{3} & \qw & \qw & \qw & \ctrl{3} & \qw & \qw & \qw & \qw & \qw & \qw & \qw & \qw & \ctrl{1} & \qw \\
& \ctrl{1} & \qw &&&& \qw & \qw & \qw & \qw & \qw & \qw & \qw & \ctrl{2} & \qw & \qw & \qw & \ctrl{2} & \qw & \qw & \qw & \qw & \qw & \qw & \ctrl{1} & \qw \\
& \ctrl{1} & \qw &&& \raisebox{0.4cm}{=} \hspace{0.9cm} & \qw & \qw & \ctrl{1} & \qw & \qw & \qw & \qw & \qw & \qw & \qw & \qw & \qw & \qw & \qw & \qw & \ctrl{1} & \qw & \qw & \ctrlo{1} & \qw \\
& \gate{Y} \qw& \qw &&&& \gate{H} & \gate{T} & \targ & \gate{T^\dag} & \gate{H} & \targ & \gate{T} & \targ & \gate{T^\dag} & \targ & \gate{T} & \targ & \gate{T^\dag} & \gate{H} & \gate{T} & \targ & \gate{T^\dag} & \gate{H} & \gate{-iZ} & \qw \\
}
$$
\end{widetext}
Here $T = |0\rangle\langle 0| + e^{i\pi/4}|1\rangle\langle 1|$. The rightmost controlled gate does not need to be applied. It would cancel with its conjugate given by $\Lambda_3(Y^\dag)$. Note that the controlled qubit needs to be connected with all three target qubits. Qiskit reports the circuit depth 34 for the above realization of $\Lambda_4(X)$ (assuming the all-to-all connectivity). If we do not apply the relative-phase Toffoli gate, then the circuit depth becomes 53. If we simply apply the built-in $\Lambda_4(X)$ gate provided in Qiskit, the circuit depth is 65, which is almost double the depth compared to our implementations. 

It is well known that the three-qubit Toffoli gate can be decomposed into single- and two-qubit gates \cite{barencoElementaryGatesQuantum1995}. The optimal realization requires six CNOT gates (with seven $T$ gates), in which all the three qubits need to be connected. More two-qubit gates are required if we have the linear connected three qubits. After optimizations, the realization of Toffoli gate on the linear connected qubits demands eight CNOT gates (with seven $T$ gates) \cite{gwinnerBenchmarking16elementQuantum2021}.

\section{\label{App:setup} Implementation setups}

\begin{figure}[t!]
\includegraphics[width=0.45\columnwidth]{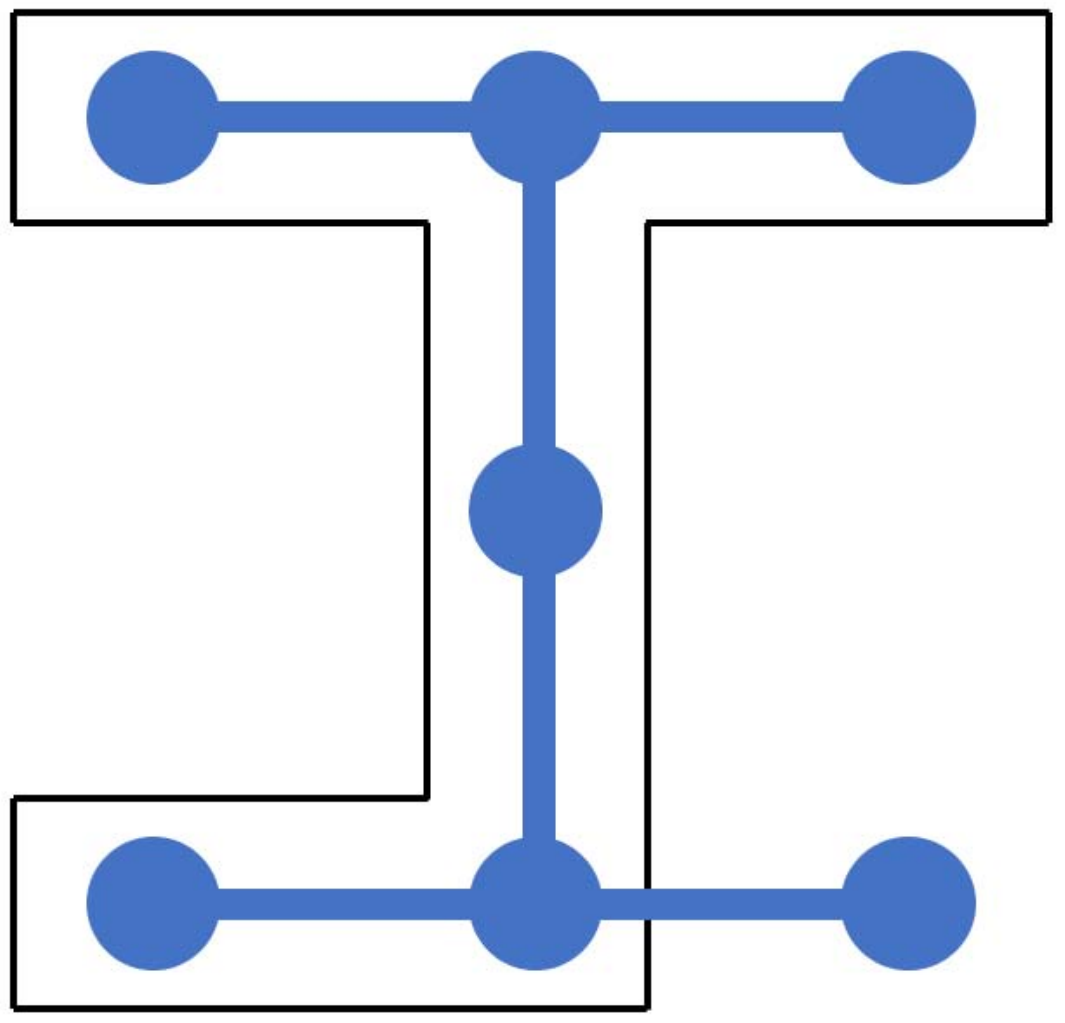}
\caption{Qubits layout of IBMQ Lagos. The boxed six dots are the physical qubits used in the implementations. }
\label{fig_layout}
\end{figure}

Since we need one ancillary qubit in the realization of five-qubit Toffoli gate $\Lambda_4(X)$, we need total of six qubits for the search circuits. IBM Quantum processors do not have full connected qubits. For example, the qubits layout of Lagos (the seven-qubit quantum computer) is shown in Fig. \ref{fig_layout}. We choose the boxed six qubits, as a ``T'' topology, in the implementations. The excluded qubit has the lowest fidelity of two-qubit gates. There is one qubit that has connections to the other three qubits. The most efficient mapping (in order to have the least number of SWAP gates) is to choose this center qubit as the ancillary. The same setups are applied to IBMQ Mumbai. The qubits in IonQ and Honeywell quantum devices are all connected. Therefore we can apply arbitrary mapping between the circuit qubits and the physical qubits. 

Although all processors support the mid-circuit measurements, we separately benchmark the two circuits in the two-stage search algorithm. It may involve additional errors if we recycle the qubits after the first stage circuit \cite{HowMeasureReset2021}. We exclude such errors in our benchmarks. Besides, we do not need to separately benchmark the circuit R2G3M3 (R3G2M2) and the second stage circuit of D2M2$\mid$D3M3 (D3M3$\mid$D2M2), since they are essentially the same. Recall that circuits R2G3M3 and R3G2M2 have random guesses on two and three bits. In order to exclude the fluctuations about the classical random guesses, we assume that we have guessed the correct target bits, and then multiply the measured success probability by $0.25$ or $0.125$ accordingly.

%\bibliographystyle{apsrev4-2}
%\bibliography{ref}

%apsrev4-2.bst 2019-01-14 (MD) hand-edited version of apsrev4-1.bst
%Control: key (0)
%Control: author (8) initials jnrlst
%Control: editor formatted (1) identically to author
%Control: production of article title (0) allowed
%Control: page (0) single
%Control: year (1) truncated
%Control: production of eprint (0) enabled
\providecommand{\noopsort}[1]{}\providecommand{\singleletter}[1]{#1}%

\end{document}